\documentclass[11pt]{article}
\pdfoutput=1
\usepackage{fullpage}
\usepackage[utf8]{inputenc}
\usepackage{cite}
\usepackage{graphicx}
\usepackage{amsmath}
\usepackage{amssymb}
\usepackage{slashed}
\title{}
\date{}

\def\beq{\begin{equation}}
\def\eeq{\end{equation}}
\def\h{\mathfrak{h}}
\begin{document}
\bibliographystyle{utphys}
\newcommand{\msbar}{\ensuremath{\overline{\text{MS}}}}
\newcommand{\DIS}{\ensuremath{\text{DIS}}}
\newcommand{\abar}{\ensuremath{\bar{\alpha}_S}}
\newcommand{\bb}{\ensuremath{\bar{\beta}_0}}
\newcommand{\rc}{\ensuremath{r_{\text{cut}}}}
\newcommand{\Nd}{\ensuremath{N_{\text{d.o.f.}}}}
\setlength{\parindent}{0pt}
\def\dd{d\!\!{}^-\!}
\def\del{\delta\!\!\!{}^-\!}

\titlepage
\begin{flushright}
Edinburgh 2016/17 \\
QMUL-PH-16-20
\end{flushright}

\vspace*{0.5cm}

\begin{center}
{\bf \Large Perturbative spacetimes from Yang-Mills theory \Large}

\vspace*{1cm}
\textsc{Andr\'{e}s Luna$^a$\footnote{a.luna-godoy.1@research.gla.ac.uk}, 
Ricardo Monteiro$^b$\footnote{ricardo.monteiro@cern.ch}, 
Isobel Nicholson$^c$\footnote{i.nicholson@sms.ed.ac.uk},\\
Alexander Ochirov$^c$\footnote{alexander.ochirov@ed.ac.uk},
Donal O'Connell$^{c}$\footnote{donal@staffmail.ed.ac.uk}, 
Niclas Westerberg$^{d,c}$\footnote{nkw2@hw.ac.uk} \\ and
Chris D. White$^e$\footnote{christopher.white@qmul.ac.uk} } \\

\vspace*{0.5cm} $^a$ School of Physics and Astronomy, University of Glasgow,\\ Glasgow G12 8QQ, Scotland, UK\\

\vspace*{0.5cm} $^b$ Theoretical Physics Department, CERN, Geneva, Switzerland\\

\vspace*{0.5cm} $^c$ Higgs Centre for Theoretical Physics, \\School of Physics and Astronomy, The University of Edinburgh,\\
Edinburgh EH9 3JZ, Scotland, UK\\

\vspace*{0.5cm} $^d$ Institute of Photonics and Quantum Sciences, School of Engineering and Physical Sciences, Heriot-Watt University, Edinburgh, UK \\

\vspace*{0.5cm} $^e$ Centre for Research in String Theory, School of Physics and Astronomy, \\
Queen Mary University of London, 327 Mile End Road, London E1 4NS, UK \\

\end{center}

\vspace*{0.5cm}

\begin{abstract}
The double copy relates scattering amplitudes in gauge and gravity
theories. In this paper, we expand the scope of the double copy to
construct spacetime metrics through a systematic perturbative expansion.
The perturbative procedure is
based on direct calculation in Yang-Mills theory, followed by
squaring the numerator of certain perturbative diagrams as specified by the double-copy algorithm.
The simplest spherically symmetric, stationary spacetime from the point of view of this procedure is
a particular member of the Janis-Newman-Winicour family of naked singularities. 
Our work paves the way for applications of the double copy to
physically interesting problems such as perturbative black-hole scattering.
\end{abstract}

\vspace*{0.5cm}

\section{Introduction}
\label{sec:intro}
Non-abelian gauge and gravity theories describe very different
physics. The former govern much of high energy physics, including
applications to particle colliders. The latter underpin most of
astrophysics and cosmology. In both types of theory, the ever
advancing experimental frontier demands theoretical precision,
including the development of new computational techniques. Recently,
an intriguing new relationship between scattering amplitudes in gauge
and gravity theories has been discovered by Bern, Carrasco and Johansson 
(BCJ)~\cite{Bern:2008qj,Bern:2010ue,Bern:2010yg}. There are
two elements in the BCJ story. The first is the {\it colour-kinematics duality},
which is the statement that it is possible to organise the numerators of
perturbative Feynman-like diagrams so that the kinematic numerator of
a given diagram obeys the same algebraic relations as the colour factor of that
diagram (for an arbitrary choice of gauge group). These relations include
Jacobi relations, which lead to three-term identities connecting planar and non-planar
diagrams in gauge theory. Furthermore, the presence of Jacobi relations 
for kinematic objects hints at the existence of an algebraic structure underlying
the gauge theory~\cite{Monteiro:2011pc}. \\

The second major element of the BCJ story is the  {\it double
copy}~\cite{Bern:2008qj,Bern:2010ue,Bern:2010yg}. This states that gauge theory
amplitudes can be straightforwardly modified to yield gravity amplitudes, essentially
by replacing the colour factor of the gauge amplitude with a second copy of the
kinematic numerator. At tree level, both the colour-kinematics duality and the double copy are proven to be
valid~\cite{BjerrumBohr:2009rd,Stieberger:2009hq,Bern:2010yg,BjerrumBohr:2010zs,Feng:2010my,Tye:2010dd,Mafra:2011kj,Monteiro:2011pc,BjerrumBohr:2012mg},
and the latter is known to be equivalent to the celebrated KLT
relations~\cite{Kawai:1985xq}, derived from string theory. However,
the BCJ story is remarkable in that it also appears to apply at
loop level, and in different types of
theory~\cite{Bern:2010ue,Bern:1998ug,Green:1982sw,Bern:1997nh,Carrasco:2011mn,Carrasco:2012ca,Mafra:2012kh,Boels:2013bi,Bjerrum-Bohr:2013iza,Bern:2013yya,Bern:2013qca,Nohle:2013bfa,Bern:2013uka,Naculich:2013xa,Johansson:2014zca,Chiodaroli:2014xia,Bern:2014sna,Mafra:2014gja,Mafra:2015mja,Johansson:2015oia,He:2015wgf,Bern:2015ooa,Mogull:2015adi,Chiodaroli:2015rdg,Chiodaroli:2015wal,Oxburgh:2012zr,Saotome:2012vy,Fu:2012uy,Fu:2013qna,Johansson:2013nsa,Cachazo:2013iea,Monteiro:2013rya,Naculich:2014rta,Naculich:2014naa,Cachazo:2014xea,Carrasco:2013ypa,Litsey:2013jfa,Nagy:2014jza,Anastasiou:2015vba,Lee:2015upy,Bjerrum-Bohr:2016axv,Carrasco:2016ldy}.\\

The existence of the double copy hints at a profound relationship
between gauge and gravity theories, that should transcend perturbative
amplitudes. To this end,
refs.~\cite{Monteiro:2014cda,Luna:2015paa,Luna:2016due,Ridgway:2015fdl} have generalised the notion of the double
copy to exact classical solutions. That is, a large family of
gravitational solutions was found that could be meaningfully
associated with a gauge theory solution, such that the relationship
between them was consistent with the BCJ double copy. These solutions
all had the special property that they linearised the Einstein and
Yang-Mills equations, so that the graviton and gauge field terminate
at first order in the coupling constant, with no higher-order
corrections. A special choice of coordinates ({\it Kerr-Schild
  coordinates}) must be chosen in the gravity theory, reminiscent of
the fact that the amplitude double copy is not manifest in all gauge
choices. An alternative approach exists, in a wide variety of
linearised supersymmetric theories, of writing the graviton as a
direct convolution of gauge
fields~\cite{Borsten:2013bp,Anastasiou:2013hba,Anastasiou:2014qba,Anastasiou:2016csv,Cardoso:2016ngt,Cardoso:2016amd}. This
in principle works for general gauge choices, but it is not yet clear
how to generalise this prescription to include nonlinear effects. One
may also consider whether the double copy can be generalised to intrinsically non-perturbative solutions, and first steps have been taken in
ref.~\cite{White:2016jzc}.\\

As is hopefully clear from the above discussion, it is not yet known
how to formulate the double copy for arbitrary field solutions, and in
particular for those which are nonlinear. However, such a procedure
would have highly useful applications. Firstly, the calculation of
metric perturbations in classical general relativity is crucial for a
plethora of astrophysical applications, but is often cumbersome. A
nonlinear double copy would allow one to calculate gauge fields
relatively simply, before porting the results to gravity. Secondly,
ref.~\cite{Luna:2015paa} provided hints that the double copy may work
in a non-Minkowski spacetime. This opens up the possibility to obtain
new insights (and possible calculational techniques) in cosmology. \\

The aim of this paper is to demonstrate explicitly how the BCJ double
copy can be used to generate nonlinear gravitational solutions
order-by-order in perturbation theory\footnote{This is the post-Minkowskian expansion, as opposed to the post-Newtonian expansion where the non-relativistic limit is also taken.}, from simpler gauge theory
counterparts. This is similar in spirit to
refs.~\cite{Bjerrum-Bohr:2013bxa,Bjerrum-Bohr:2014zsa,Bjerrum-Bohr:2016hpa},
which extracted both classical and quantum gravitational corrections
from amplitudes obtained from gauge theory ingredients; and to refs.~\cite{Duff:1973zz,Neill:2013wsa}, which used tree-level amplitudes to construct perturbatively the Schwarszchild spacetime. Very recently, ref.~\cite{Goldberger:2016iau} has studied the double copy procedure for classical radiation emitted by multiple point charges. Here we take a
more direct approach, namely to calculate the graviton field
generated by a given source, rather than extracting this
from a scattering amplitude. Another recent work, ref.~\cite{Chu:2016ngc}, proposes applications to cosmological gravitational waves, pointing out a double copy of radiation memory. \\

As will be explained in detail in what follows, our scheme involves
solving the Yang-Mills equations for a given source order-by-order in
the coupling constant. We then copy this solution by duplicating
kinematic numerators, before identifying a certain product of gauge
fields with a two-index field $H_{\mu\nu}$, motivated by
ref.~\cite{Bern:2010yg}. This field contains degrees of freedom
associated with a conventional graviton $h_{\mu\nu}$, together with a
scalar field $\phi$ and two-form field $B_{\mu\nu}$. For convenience,
we will refer to $H_{\mu\nu}$ as the {\it fat graviton}, and the
physical field $h_{\mu\nu}$ as the {\it skinny graviton}. As we will
see, the skinny fields $h_{\mu\nu}$, $B_{\mu\nu}$ and $\phi$ can
be obtained from knowledge of $H_{\mu\nu}$, though this extraction
requires knowledge of a certain
gauge transformation and field redefinition in general.\\

The structure of our paper is as follows. In section~\ref{sec:review},
we briefly review the BCJ double copy. In section~\ref{sec:linear}, we
work at leading order in perturbation theory, and outline our
procedure for obtaining gravity solutions from Yang-Mills fields. In
section~\ref{sec:nonlin}, we work to first and second subleading order in
perturbation theory, thus explicitly demonstrating how nonlinear
solutions can be generated in our approach. Finally, we discuss our
results and conclude in section~\ref{sec:discuss}.

\section{Review of the BCJ double copy}
\label{sec:review}

Our aim in this section is to recall salient details about the BCJ
double copy~\cite{Bern:2008qj,Bern:2010ue,Bern:2010yg}, that will be needed in
what follows. Since we will be dealing with solutions to the classical theories, we are only concerned with the tree-level story, which is well established, whereas at loop level the BCJ proposal is a conjecture.
First, we recall that an $m$-point tree-level amplitude
in non-abelian gauge theory may be written in the general form
\begin{equation}
{\cal A}_m=g^{m-2}\sum_{i\in\Gamma} \frac{n_i\,c_i}{\prod_{\alpha_i}p_{\alpha_i}^2},
\label{ampform}
\end{equation}
where $g$ is the coupling constant, and the sum is over the set of cubic graphs
$\Gamma$. The denominator arises from propagators associated with each internal
line, and $c_i$ is a colour factor obtained by dressing each vertex
with structure constants. Finally, $n_i$ is a kinematic numerator,
composed of momenta and polarisation vectors. Note that the sum over
graphs involves cubic topologies only, despite the fact that
non-abelian gauge theories include quartic interaction terms for the
gluon. These can always be broken up into cubic-type graph contributions, so that
eq.~(\ref{ampform}) is indeed fully general. The form is not unique,
however, owing to the fact that the numerators $\{n_i\}$ are modified
by gauge transformations and / or field redefinitions, neither of
which affect the amplitude. A compact way to summarise this is that
one is free to modify each individual numerator according to the
{\it generalised gauge transformation}
\begin{equation}
n_i\rightarrow n_i+\Delta_i, \qquad \quad \sum_i\frac{\Delta_i c_i}
{\prod_{\alpha_i}p_{\alpha_i}^2}=0,
\label{gengauge}
\end{equation}
where the latter condition expresses the invariance of the amplitude. \\

The set of cubic graphs in
eq.~(\ref{ampform}) may be divided into overlapping sets of three,
where the colour factors $c_i$ are related by Jacobi identities, associated to the Lie algebra of the colour group. Remarkably, it is possible to
choose the numerators $n_i$ so that they obey similar Jacobi
identities, which take the form of coupled functional equations. This
property is known as {\it colour-kinematics duality}, and hints at an intriguing
correspondence between colour and kinematic degrees of freedom that is
still not fully understood, although progress has been made in the
self-dual sector of the theory~\cite{Monteiro:2011pc}. More generally, the field-theory limit of superstring theory has been very fruitful for understanding
colour-kinematics duality~\cite{Mafra:2011kj,Lee:2015upy,Mafra:2015vca} and there has been recent progress on more formal aspects of the duality~\cite{Fu:2016plh,Brown:2016mrh,Brown:2016hck}. \\

Given a gauge theory amplitude in BCJ-dual form, the {\it double copy}
prescription states that
\begin{equation}
{\cal M}_m=i\left(\frac{\kappa}{2}\right)^{m-2}
\sum_{i\in\Gamma} \frac{n_i\,\tilde{n}_i}{\prod_{\alpha_i}p_{\alpha_i}^2}
\label{ampform2}
\end{equation}
is an $m$-point gravity amplitude, where 
\begin{equation}
g_{\mu\nu}=\eta_{\mu\nu}+\kappa h_{\mu\nu}
\label{hdef}
\end{equation}
can be chosen to define the graviton field,
and $\kappa=\sqrt{32\pi G}$ is the gravitational coupling constant.\footnote{We work in the mostly plus metric convention.}
This result is obtained from eq.~(\ref{ampform}) by replacing the gauge theory
coupling constant with its gravitational counterpart, and colour
factors with a second set of kinematic numerators $\tilde{n}_i$. Therefore, the procedure modifies
the numerators of amplitudes term by term, but leaves the denominators
in eqs.~(\ref{ampform}, \ref{ampform2}) intact. A similar phenomenon
occurs in the double copy for exact classical solutions of
refs.~\cite{Monteiro:2014cda,Luna:2015paa,Luna:2016due}, in which
scalar propagators play a crucial role. \\

The gravity theory associated with the scattering amplitudes \eqref{ampform2} depends on the two gauge
theories from which the numerators $\{n_i\}$, $\{\tilde{n}_i\}$ are
taken. In this paper, both will be taken from pure Yang-Mills theory,
which is mapped by the double copy to ``${\cal N}=0$ supergravity''. This
theory is defined as Einstein gravity coupled to a scalar field $\phi$ (known as the dilaton) and a two-form $B_{\mu\nu}$ (known as the Kalb-Ramond field, which can be replaced by an axion in four spacetime dimensions). The action for these fields is
\begin{equation}
S=\int d^D x\sqrt{-g}\left[\frac{2}{\kappa^2}R
-\frac{1}{2(D-2)}\partial^\mu\phi \partial_\mu \phi
-\frac{1}{6}e^{-2\kappa\phi/{D-2}}H^{\lambda\mu\nu}
H_{\lambda\mu\nu}\right],
\label{SN=0}
\end{equation}
where $H_{\lambda\mu\nu}$ is the field strength of $B_{\mu\nu}$.
In the following, we will study perturbative solutions of this theory around Minkowski space. The starting point is to consider linearised fields, for which the equations of motion are
\begin{align}
\partial^2 h_{\mu\nu} &- \partial_\mu \partial^\rho h_{\rho \nu} - \partial_\nu \partial^\rho h_{\rho \mu} + \partial_\mu \partial_\nu h + \eta_{\mu\nu} \left[\partial^\rho \partial^\sigma h_{\rho\sigma} - \partial^2 h \right]  = 0,\notag \\
\partial^2 B_{\mu\nu} &- \partial_\mu \partial^\rho B_{\rho \nu} + \partial_\nu \partial^\rho B_{\rho \mu}  = 0 , \notag \\ 
\partial^2 \phi & = 0.
\label{skinnyEOM}
\end{align}

Instead of the straightforward graviton field $h_{\mu\nu}$ defined by \eqref{hdef}, we will often work with the ``gothic'' metric perturbation $\h^{\mu\nu}$ such that
\begin{equation}
\sqrt{-g}\,g^{\mu\nu} = \eta^{\mu\nu} - \kappa\, \h^{\mu\nu},
\label{gothich}
\end{equation}
as it is common in perturbation theory \cite{poisson2014gravity}. In terms of this gothic graviton field, the de Donder gauge condition is simply $\partial_\mu \h^{\mu\nu} = 0$ to all orders. At the linear order, the two metric perturbations are simply related:
\begin{equation}
\h_{\mu\nu} = h_{\mu\nu} - \frac12 \eta_{\mu\nu} h,
\end{equation}
and the linear gauge transformation generated by $x^\mu \to x^\mu -\kappa\, \xi^\mu$ is
\begin{equation}
\h_{\mu\nu} \; \to \; \h'_{\mu\nu} = \h_{\mu\nu} + \partial_\mu \xi_\nu + \partial_\nu \xi_\mu - \eta_{\mu\nu} \partial \cdot \xi.
\end{equation}
This transformation is more convenient in what follows than the standard gauge transformation for $h_{\mu\nu}$ (where the last term is missing). Finally, the linearised equation of motion is
\begin{equation}
\partial^2 \h_{\mu\nu} - \partial_\mu \partial^\rho \h_{\rho \nu} - \partial_\nu \partial^\rho \h_{\rho \mu} + \eta_{\mu\nu} \partial^\rho \partial^\sigma \h_{\rho\sigma} = 0.
\end{equation}
In de Donder gauge, we have simply $\partial^2 \h_{\mu\nu}=0$.

\section{Linear gravitons from Yang-Mills fields}
\label{sec:linear}

Our goal is to rewrite gravitational perturbation theory in terms of the fat graviton $H_{\mu\nu}$, rather than more standard perturbative fields such as $\{\mathfrak{h}_{\mu\nu}, B_{\mu\nu}, \phi \}$. The idea is that the fat graviton is the field whose interactions are directly dictated by the double copy from gauge theory. In this section, we will discuss in some detail the mapping between the skinny fields and the fat graviton at the linearised level. Indeed, we will see that there is an invertible map, so that the fat graviton may be constructed from skinny fields $H_{\mu\nu} = H_{\mu\nu} (\mathfrak{h}_{\alpha\beta}, B_{\alpha\beta}, \phi)$, but also the skinny fields can be determined from the fat field, $\mathfrak{h}_{\mu\nu} = \mathfrak{h}_{\mu\nu}(H_{\alpha\beta}), B_{\mu\nu} = B_{\mu\nu}(H_{\alpha\beta}), \phi = \phi(H_{\alpha \beta})$. We will determine the relations between the fields beginning with the simplest case: linearised waves.

\subsection{Linear waves}
As a prelude to obtaining non-linear gravitational solutions from
Yang-Mills theory, we first discuss linear solutions of both
theories. The simplest possible solutions are linear waves. These are
well-known to double copy between gauge and gravity theories (see
e.g.~\cite{Siegel:1999ew}). This property is crucial
for the double copy description of scattering amplitudes, whose incoming and
outgoing states are plane waves. Here, we use linear waves to motivate
a prescribed relationship between fat and skinny fields, which will be
generalised in later sections.\\

Let us start by considering a gravitational plane wave in the de
Donder gauge. The free equation of motion for the graviton is simply $\partial^2 \h_{\mu\nu} = 0$. Plane wave solutions take the form
\begin{equation}
\h_{\mu\nu}=a_{\mu\nu}e^{ip\cdot x}, \qquad \quad
p^\mu a_{\mu\nu}=0, \qquad \quad
p^2=0,
\label{planewaveDD}
\end{equation}
where $a_{\mu\nu}$ is a constant tensor, and the last condition
follows from the equation of motion. Symmetry of the graviton implies
$a_{\mu\nu}=a_{\nu\mu}$, and one may also fix a residual gauge freedom
by setting $a\equiv a^\mu_\mu=0$, so that $\h_{\mu\nu}$ becomes a traceless, symmetric matrix. 
It is useful to further characterise the
matrix $a_{\mu\nu}$ by introducing a set of $(D-2)$ polarisation
vectors $\epsilon^i_\mu$ satisfying the orthogonality conditions
\begin{equation}
p\cdot \epsilon^i=0, \qquad \quad
q\cdot \epsilon^i=0,
\label{orthogonality}
\end{equation}
where $q^\mu$ ($q^2=0$, $p\cdot q\neq 0$) is an auxiliary null vector
used to project out physical degrees of freedom for an on-shell
massless vector boson. These polarisation vectors are a complete set, so they satisfy a 
completeness relation
\begin{equation}
\epsilon^i_\mu \epsilon^i_\nu=\eta_{\mu\nu}-\frac{p_\mu q_\nu+p_\nu q_\mu}{p\cdot q}.
\label{completeness}
\end{equation}
Then the equation of motion for $\h_{\mu\nu}$,
together with the symmetry and gauge conditions on $a_{\mu\nu}$, imply
that one may write
\begin{equation}
a_{\mu\nu} = f^\slashed{t}_{ij} \epsilon^i_\mu \epsilon^j_\nu ,
\end{equation}
where $f^\slashed{t}_{ij}$ is a traceless symmetric matrix. Thus, the
linearised gravitational waves have polarisation states which can be
constructed from outer products of vector waves, times traceless
symmetric matrices. \\

Similarly, one may consider linear plane wave solutions for a two-form
and $\phi$ field. Imposing Lorenz gauge $\partial^\mu B_{\mu\nu} = 0$ for
the antisymmetric tensor, its free equation of motion becomes simply $\partial^2 B_{\mu\nu} = 0$.
Thus plane wave solutions are
\begin{equation}
B_{\mu\nu} = \tilde f_{ij} \epsilon^i_\mu \epsilon^j_\nu e^{i p \cdot x},
\end{equation}
where $\tilde f_{ij}$ is a constant antisymmetric matrix. Meanwhile the free equation of motion for the scalar
field is $\partial^2 \phi = 0$, with plane wave solution
\begin{equation}
\phi = f_\phi e^{i p \cdot x}.
\end{equation}
The double copy associates these skinny waves with a single fat
graviton field $H_{\mu\nu}$ satisfying the field equation $\partial^2H_{\mu\nu}=0$, 
\begin{equation}
H_{\mu\nu}=f_{ij}\epsilon^i_\mu \epsilon^j_\nu e^{ip\cdot x},
\label{Hsol}
\end{equation}
where now $f_{ij}$ is a general $D-2$ matrix and we have chosen a gauge condition $\partial^\mu H_{\mu\nu} = 0 = \partial^\mu H_{\nu\mu}$. One may write this
decomposition as
\begin{align}
H_{\mu\nu} &= \left(f^\slashed{t}_{ij} + \tilde f_{ij} + \delta_{ij}
\frac{f_\phi}{D-2} \right) \epsilon^i_\mu \epsilon^j_\nu  e^{ip\cdot x}, \\
&= \h_{\mu\nu} + B_{\mu\nu} + \left( \eta_{\mu\nu} -
\frac{p_\mu q_\nu + p_\nu q_\mu}{p \cdot q} \right) \frac{\phi}{D-2},
\label{Hdecomp}
\end{align}
which explicitly constructs the fat graviton from skinny fields. Working in position space for constant $q$, this becomes
\begin{equation}
H_{\mu\nu}(x) = \h_{\mu\nu}(x) + B_{\mu\nu}(x) + P^q_{\mu\nu} \phi,
\label{HdecompPosn}
\end{equation}
where we have defined the projection operator
\begin{equation}
P^q_{\mu\nu} = \frac{1}{D-2}\left( \eta_{\mu\nu} - \frac{q_\mu \partial_\nu + q_\nu \partial_\mu}{q \cdot \partial} \right),
\label{projectorq}
\end{equation}
which will be important throughout this article.\footnote{Notice that $\hat{P}^q_{\mu\nu}=(D-2)P^q_{\mu\nu}$ is the properly normalised projection operator, such that $\hat{P}^{q\,\lambda}_\mu \hat{P}^{q\,\nu}_\lambda=\hat{P}^{q\,\nu}_\mu$, and $\hat{P}^{q\,\mu}_\mu=D-2$.} \\

Our goal in this work is not to construct fat gravitons from skinny fields, but on the contrary to determine skinny fields using a perturbative expansion based on the double copy and the fat graviton. Therefore it is important that we can determine the skinny fields given knowledge of the fat graviton. To that end, recall that we have been able to choose a gauge so that the trace, $\h$, of the metric perturbation vanishes. Therefore the trace of the fat graviton determines the dilaton:
\begin{equation}
\phi = H^\mu{}_\mu \equiv H.
\end{equation}
We may now use symmetry to determine the skinny graviton and antisymmetric tensor from the fat graviton:
\begin{align}
B_{\mu\nu} &= \frac12 \left( H_{\mu\nu} - H_{\nu\mu} \right), \\
\h_{\mu\nu} &= \frac12 \left( H_{\mu\nu} + H_{\nu\mu} \right) - P^q_{\mu\nu} H.
\end{align}
The basic strategy of this construction is simple: we have decomposed the matrix field $H_{\mu\nu}$ into its antisymmetric, traceless symmetric, and trace parts. \\

It is worth dwelling on the decomposition of the fat graviton into skinny fields a little further. Having constructed $\h_{\mu\nu}$ from the fat graviton, we are free to consider a gauge transformation of the skinny graviton:
\begin{align}
\h'_{\mu\nu} &= \h_{\mu\nu} + \partial_\mu \xi_\nu + \partial_\nu \xi_\mu - \eta_{\mu\nu} \partial \cdot \xi \\
&= \frac12 \left( H_{\mu\nu} + H_{\nu\mu} \right) - \frac{1}{D-2}\left( \eta_{\mu\nu} - \frac{q_\mu \partial_\nu + q_\nu \partial_\mu}{q \cdot \partial} \right) H + \partial_\mu \xi_\nu + \partial_\nu \xi_\mu - \eta_{\mu\nu} \partial \cdot \xi.
\end{align}
If we choose 
\begin{equation}
\xi_\mu = -  \frac{1}{D-2} \left( \frac{q_\mu}{q \cdot \partial} \right) H,
\end{equation}
then we find that the expression for the $\h'_{\mu\nu}$ simplifes to
\begin{equation}
\h'_{\mu\nu} = \frac12 \left( H_{\mu\nu} + H_{\nu\mu} \right).
\end{equation}
Thus, up to a gauge transformation, the skinny graviton is the symmetric part of the fat graviton. It may be worth emphasising that
$\phi$ and $B_{\mu\nu}$ also transform under this gauge transformation, which is, of course, a particular diffeomorphism. However, the transformation of $\phi$ and $B_{\mu\nu}$ is suppressed by a power of $\kappa$, and so we may take them to be gauge invariant for diffeomorphisms at this order.\\

We will see below that the perturbative expansion for fat gravitons is much simpler than the perturbative expansion for the individual skinny fields. But before we embark on that story, it is important to expand our understanding of the relationship between the fat graviton and the skinny fields beyond the sole case of plane waves.

\subsection{General linearised vacuum solutions}

For plane waves, the fat graviton is given in terms of skinny fields
in eq.~(\ref{HdecompPosn}), and at first glance this equation is not
surprising: one may always choose to decompose an arbitrary rank two
tensor into its symmetric traceless, antisymmetric and trace
parts. However, eq.~(\ref{HdecompPosn}) contains non-trivial physical
content, namely that the various terms on the RHS are the genuine
propagating degrees of freedom associated with each of the skinny
fields. The auxiliary vector $q_\mu$ plays a crucial role here: it is
associated in the gauge theory with the definition of physical
polarisation vectors, and thus can be used to project out physical
degrees of freedom in the gravity theory. One may then ask whether
eq.~(\ref{HdecompPosn}) generalises for arbitrary solutions of the
linearised equations of motion. There is potentially a problem in that
the relationship becomes ambiguous: the trace of the skinny graviton
may be nonzero (as is indeed the case in general gauges), and one must
then resolve how the trace degree of freedom in $H^{\mu\nu}$ enters
the trace of the skinny graviton, and the scalar field
$\phi$. Furthermore, it is not immediately clear that
eq.~(\ref{HdecompPosn}) (derived for plane waves) will work when non-zero
sources are present in the field equations. In order to use the double
copy in physically relevant applications, we must consider this
possibility.\\

Here we will restrict ourselves to skinny gravitons that are in de
Donder gauge. However, we will relax the traceless condition on the
skinny graviton which was natural in the previous section. To account
for the trace, we
postulate that eq.~(\ref{HdecompPosn}) should be replaced by
\begin{equation}
H_{\mu\nu}(x) = \h_{\mu\nu}(x) + B_{\mu\nu}(x) + P^q_{\mu\nu} (\phi - \h).
\label{eq:linearFatDef}
\end{equation}
To be useful, this definition of the fat graviton must be invertible. First, note that the trace of $H_{\mu\nu}$ determines $\phi$ as before, while the antisymmetric part of $H_{\mu\nu}$ determines $B_{\mu\nu}$. Finally, the traceless symmetric part of the fat graviton is
\begin{equation}
\frac12\left(H_{\mu\nu} + H_{\nu\mu}  \right) - P^q_{\mu\nu} H = \h_{\mu\nu}(x) - P^q_{\mu\nu} \h = \h'_{\mu\nu}(x),
\label{guts}
\end{equation}
where $\h'_{\mu\nu}(x)$ is a gauge transformation of $\h_{\mu\nu}(x)$. In practice, we find it useful to work with $\h_{\mu\nu}(x)$ rather than $\h'_{\mu\nu}(x)$, because at higher orders the gauge transformation to $\h'_{\mu\nu}(x)$ leads to more cumbersome formulae. 
It is also worth noticing that both $\h_{\mu\nu}$ and $\h'_{\mu\nu}$ are in de Donder gauge, since
\begin{equation}
\partial^\mu P^q_{\mu\nu} \h = \frac{1}{D-2}\left( \partial_\nu - \frac{q_\nu \partial^2  + q\cdot\partial \, \partial_\nu}{q \cdot \partial} \right) \h 
= - \frac{1}{D-2}\frac{q_\nu}{q \cdot \partial} \partial^2 \h 
= 0.
\end{equation}

Our relationship between skinny and fat fields still holds only for linearised fields; we
will explicitly find corrections to eq.~(\ref{eq:linearFatDef}) at
higher orders in perturbation theory in
section~\ref{sec:nonlin}. Before doing so, however, it is instructive to
illustrate the above general discussion with some specific solutions
of the linear field equations, showing how the fat and skinny fields
are mutually related. \\

\subsection{The linear fat graviton for Schwarzschild}
\label{sec:fatlinearSchwarzschild}

One aim of our programme is to be able to describe scattering
processes involving black holes. To this end, let us see how to extend
the above results in the presence of point-like masses. It is easy to construct a fat graviton for the linearised Schwarzschild metric: we begin by noticing that, in the case of Schwarzschild
($D=4$), we have
\begin{equation}
\h_{\mu\nu}(r) =  \frac{\kappa}{2} \frac{M}{4\pi r} u_\mu u_\nu 
+{\cal O}(\kappa^2) , \qquad
B_{\mu\nu}(x) = 0 , \qquad
\phi(x) = 0 , \qquad \text{with}\;\; u_\mu=(1,0,0,0).
\label{Schwskinnylinear}
\end{equation}
The fat graviton depends on an arbitrary constant null vector $q^\mu$. In this section, for illustration, we will make an explicit choice of $q^\mu=(-1,0,0,1)$ and evaluate the action of the projector~\eqref{projectorq}
in position space in full. A computation gives
\begin{align}
H_{\mu\nu}&= \frac{\kappa}{2} \frac{M}{4\pi r} u_\mu u_\mu + P^q_{\mu\nu} \left( \frac{\kappa}{2} \frac{M}{4\pi r} \right) \\
&=\frac{\kappa}{2} \frac{M}{4\pi r}\left(u_\mu u_\nu + \frac{1}{2} (\eta_{\mu\nu} - q_\mu l_\nu - q_\nu l_\mu) \right),
\label{Schwfatlinear}
\end{align}
where $l_\mu=(0,x,y,r+z)/(r+z)$, such that $q\cdot l=1$.
It is easy to check that $\partial^\mu H_{\mu\nu}=0$, $\partial^2H_{\mu\nu}=0$. \\

Going in the other direction, it is easy to compute the skinny fields given this fat graviton. Since $H_{\mu\nu}$ is traceless, the dilaton vanishes. Similarly $H_{\mu\nu}$ is symmetric, and therefore $B_{\mu\nu} = 0$. The skinny graviton can therefore be taken to be equal to the fat graviton. While this result seems to be at odds with \eqref{Schwskinnylinear}, recall that they differ only by a gauge transformation (which leaves $\phi$ and $B_{\mu\nu}$ unaffected at this order) and that the skinny graviton we recover is traceless, as we would expect from eq.~\eqref{guts}.
\\

It may not seem that we have gained much by passing to eq.~\eqref{Schwfatlinear} from eq.~\eqref{Schwskinnylinear}. However, it is our contention that it is simpler to compute perturbative corrections to metrics using the formalism of the fat graviton than with the traditional approach. We will illustrate this in a specific example later in this paper.

\subsection{Solutions with linearised dilatons}
\label{sec:linearJNW}

The linearised Schwarzschild metric corresponds to a somewhat complicated fat graviton. Since the fat graviton's equation of motion is simply $\partial^2 H_{\mu\nu} = 0$, it is natural to consider the solution
\begin{equation}
\label{eq:fatJNW}
H_{\mu\nu} = \frac{\kappa}{2} \frac{M}{4\pi r} u_\mu u_\nu ,\qquad \text{with}\;\; u_\mu=(1,0,0,0),
\end{equation}
which corresponds to inserting a singularity at the origin. We will see that this solution has the physical interpretation of a point mass which is also a source for the scalar dilaton. Indeed, the dilaton contained in the fat graviton is given by its trace:
\begin{equation}
\phi = - \frac{\kappa}{2} \frac{M}{4\pi r}.
\end{equation}
Since the fat graviton is symmetric, $B_{\mu\nu} = 0$. Meanwhile the skinny graviton is
\begin{equation}
\h_{\mu\nu} = \frac{\kappa}{2} \frac{M}{4\pi r}\left(u_\mu u_\nu + \frac{1}{2} (\eta_{\mu\nu} - q_\mu l_\nu - q_\nu l_\mu) \right).
\end{equation}
Again, a linearised diffeomorphism can give the skinny graviton the same form as the fat graviton. \\

It is natural to ask what is the non-perturbative static spherically-symmetric solution for which we are finding the linearised fields. Exact solutions of the Einstein equations minimally coupled to a scalar field of this form were discussed by Janis, Newman and Winicour (JNW)~\cite{Janis:1968zz} and have been extensively studied in the literature~\cite{Fisher:1948yn,Buchdahl:1959nk,Janis:1968zz,Bronnikov:1973fh,Wyman:1981bd,Virbhadra:1997ie,Bhadra:2001fx}. The complete solution is, in fact, a naked singularity, consistent with the no-hair theorem. The general JNW metric and dilaton can be expressed as
\begin{align}
ds^2 & = - \left(1 - \frac{\rho_0}{\rho}\right)^\gamma dt^2 +\left(1 - \frac{\rho_0}{\rho}\right)^{-\gamma} d\rho^2 + \left(1 - \frac{\rho_0}{\rho}\right)^{1-\gamma} \rho^2 d\Omega^2, \\
\phi & = \frac{\kappa}{2} \frac{Y}{4\pi \rho_0} \log \left( 1 - \frac{\rho_0}{\rho} \right).
\end{align}
where the two parameters $\rho_0$ and $\gamma$ can be given in terms of the mass $M$ and the scalar coupling $Y$ as
\begin{equation}
\rho_0 = 2 G \sqrt{M^2 + Y^2}= \left(\frac{\kappa}{2}\right)^2 \frac{\sqrt{M^2 + Y^2}}{4\pi}, \qquad
\gamma = \frac{M}{\sqrt{M^2 + Y^2}}.
\end{equation}
For $Y=0$ and $M>0$, we recover the Schwarzschild black hole, with the event horizon at $\rho=\rho_0$. For $|Y|>0$ and $M>0$, the solution also decays for large $\rho$, but there is a naked singularity at $\rho=\rho_0$, which now corresponds to zero radius (since the metric factor in front of $d\Omega^2$ vanishes). We can write the JNW solution in de Donder gauge by applying the coordinate transformation $\rho=r+\rho_0/2$, where $r$ is the Cartesian radius in the de Donder coordinates. Expanding in $\kappa$, the result is
\begin{align}
\label{eq:JNWexpansion}
\h_{\mu\nu} & = \frac{\kappa}{2} \frac{M}{4\pi r} u_\mu u_\nu +\left(\frac{\kappa}{2}\right)^3 
\frac{1}{8(4\pi r)^2}
\big((7M^2-Y^2)u_\mu u_\nu+(M^2+Y^2) \hat{r}_\mu \hat{r}_\nu\big) + \mathcal{O}(\kappa^5), \\
\phi & = -\frac{\kappa}{2} \frac{Y}{4\pi r} + \mathcal{O}(\kappa^5),
\end{align}
with $\hat{r}^\mu=(0,\mathbf{x}/r)$.
Despite its somewhat esoteric nature, this naked singularity is a particularly natural object from the point of view of the perturbative double copy. At large distances from the singularity, both the metric perturbation and the scalar field fall off as $1/r$, and for $Y=M$ this leading part reproduces the skinny fields obtained above, up to a linearised diffeomorphism in~$\h_{\mu\nu}$. In Section~\ref{sec:nonlin}, we will discuss the first two non-linear corrections to the JNW metric using fat gravitons, and, in the case of the first correction, we will match the expansion above. We conclude that the JNW solution with $Y=M$ is the exact solution associated to the linearised fat graviton~\eqref{eq:fatJNW}. \\

We can also ask what fat graviton would be associated to the general JNW family of solutions, with $M$ and $Y$ generic. Since we are dealing with linearised fields, we can superpose contributions, and so we arrive at
\begin{equation}
H_{\mu\nu}=\frac{\kappa}{2} \frac{1}{4\pi r}\left(M \,u_\mu u_\nu + (M-Y) \;\frac{1}{2} (\eta_{\mu\nu} - q_\mu l_\nu - q_\nu l_\mu) \right).
\end{equation}
The gauge theory ``single copy'' associated to this field is simply
the Coulomb solution, which presents an apparent puzzle:
ref.~\cite{Monteiro:2014cda} argued that the double copy of the
Coulomb solution is a pure Schwarzschild black hole, with no dilaton
field. Above, however, the double copy produces a JNW solution. The
latter was also found in ref.~\cite{Goldberger:2016iau}, which thus
concluded that the Schwarzschild solution is not obtained by
the double copy, but can only be true in certain limits (such as the
limit of an infinite number of dimensions). The resolution of this
apparent contradiction is that one can choose whether or not
the dilaton is sourced upon taking the double copy. It is well-known
in amplitude calculations, for example, that gluon amplitudes can
double copy to arbitrary combinations of amplitudes for gravitons,
dilatons and/or B-fields. A simple example are amplitudes for linearly polarised
gauge bosons: the double copied ``amplitude'' involves mixed waves of
gravitons and dilatons. Thus, the result in the gravity theory
depends on the linear combinations of the pairs of
gluon polarisations involved in the double copy. Here, we may say that
the Schwarzschild solution is a double copy of the Coulomb potential,
as given by the Kerr-Schild double copy \cite{Monteiro:2014cda}, just
as one may say that appropriate combinations of amplitudes of gluons
lead to amplitudes of pure gravitons. The analogue of more general gravity
amplitudes with both gravitons and dilatons, obtained via the double
copy, is the JNW solution. Therefore the double copy of the Coulomb solution
is somewhat ambiguous: in fact, it is any member of the JNW
family of singularities, including the Schwarzschild metric. Note that the Kerr-Schild
double copy is applicable only in the Schwarzschild special case since the other members
of the JNW family of spacetimes do not admit Kerr-Schild coordinates. \\


For the vacuum Kerr-Schild solutions studied in
\cite{Monteiro:2014cda}, in particular for the Schwarzschild black
hole, it was possible to give an exact map between the gauge theory
solution and the exact graviton field, making use of Kerr-Schild
coordinates (as opposed to the de Donder gauge used here). For the
general JNW solution, the double copy correspondence was inferred
above from the symmetries of the problem and from the perturbative
results. A more general double copy map would also be able to deal
with the exact JNW solution. This remains an important goal, but one which 
is not addressed in this paper.

\section{Perturbative Corrections}
\label{sec:nonlin}

Now that we have understood how to construct fat gravitons in several cases, let us finally put them to use. In this section, we will construct nonlinear perturbative correction to spacetime metrics and/or dilatons using the double copy. Thus, we will map the problem of finding perturbative corrections to a simple calculation in gauge theory. 

\subsection{Perturbative metrics from gauge theory}
\label{sec:firstNLO}

Since the basis of our calculations is the perturbative expansion of gauge theory, we begin with the vacuum Yang-Mills equation
\begin{equation}
\partial^\mu F^a_{\mu\nu} + g f^{abc} A^{b\mu} F^c_{\mu\nu} = 0,
\end{equation}
where $g$ is the coupling constant, while the field strength tensor is
\begin{equation}
F^a_{\mu\nu} = \partial_\mu A^a_\nu - \partial_\nu A^a_\mu + g f^{abc} A^b_\mu A^c_\nu.
\end{equation}
We are interested in a perturbative solution of these equations, so that the gauge field $A^a_\mu$ can be written as a power series in the coupling:
\begin{equation}
A^a_\mu = A^{(0)a}_\mu + g A^{(1)a}_\mu + g^2 A^{(2)a}_\mu + \cdots.
\end{equation}
In this expansion, the perturbative coefficients $A^{(i)a}_\mu$ are assumed to have no dependence on the coupling $g$. We use a similar notation for the perturbation series for the skinny and fat gravitons:
\begin{align}
   \h^{\mu\nu} & = \h^{(0)\mu\nu} + \frac{\kappa}{2} \h^{(1)\mu\nu}
                 + \left(\frac{\kappa}{2}\right)^{2} \h^{(2)\mu\nu}
                 + \cdots , \\
   H^{\mu\nu} & = H^{(0)\mu\nu} + \frac{\kappa}{2} H^{(1)\mu\nu} 
                 + \left(\frac{\kappa}{2}\right)^{2} H^{(2)\mu\nu} + \cdots.
\end{align}

We can construct solutions in perturbation theory in a straightforward manner. To zeroth order in the coupling, the Yang-Mills equation in Lorenz gauge $\partial^\mu A^a_\mu = 0$ is simply
\begin{equation}
\partial^2 A^{(0)a}_\mu = 0.
\end{equation}
For our present purposes, two basic solutions of this equation will be of interest: wave solutions, and Coulomb-like solutions with isolated singularities. \\

Given a solution $A^{(0)a}_\mu$ of the linearised Yang-Mills equation, it is easy to write down an expression for the first order correction $A^{(1)a}_\mu$ by expanding the Yang-Mills equation to first order in $g$:
\begin{equation}
\partial^2 A^{(1)a}_\nu  = -2 f^{abc} A^{(0)b\mu} \partial_\mu A^{(0)c}_\nu +   f^{abc} A^{(0)b\mu} \partial_\nu A^{(0)c}_\mu.
\end{equation}
The double copy is most easily understood in Fourier (momentum)
space. To simplify our notation, we define
\begin{equation}
\int \dd^D p F(p) \equiv \int \frac{d^D p}{(2\pi)^D} F(p), \qquad \quad
\del^D(p) \equiv (2\pi)^D \delta^{(D)}(p).
\end{equation}
Using this notation, we may write the solution for the first perturbative correction in Fourier space in the familiar form
\begin{multline}
A^{(1)a\mu}(-p_1) = \frac{i}{2p_1^2} f^{abc} \int \dd^D p_2 \dd^D p_3 \del^D(p_1 + p_2 + p_3) \\
\times \left[  (p_1 - p_2)^\gamma \eta^{\mu\beta} + (p_2 - p_3)^\mu \eta^{\beta\gamma} + (p_3 - p_1)^\beta \eta^{\gamma\mu} \right]  A^{(0)b}_\beta(p_2) A^{(0)c}_\gamma(p_3) .
\label{eq:correctionYM}
\end{multline}
Notice that the factor in square brackets in this equation obeys the same algebraic symmetries as the colour factor, $f^{abc}$, appearing in the equation. This is a requirement of colour-kinematics duality. Before using the double copy, it is necessary to ensure that this duality holds. \\

The power of the double copy is that it is now completely trivial to compute the perturbative correction $H^{(1)}_{\mu\nu}$ to a linearised fat graviton $H^{(0)}_{\mu\nu}$. All we need to do, following \cite{Bern:2008qj,Bern:2010ue,Bern:2010yg}, is to square the numerator in eq.~\eqref{eq:correctionYM}, ignore the colour structure, and assemble fat gravitons by the rule that $A^{(0)a}_\mu(p) A^{(0)b}_\nu(p) \rightarrow H^{(0)}_{\mu\nu}(p)$. This straightforward procedure leads to
\begin{align}\!\!\!
H^{(1)\mu\mu'}(-p_1) = \frac{1}{4p_1^2} & \int \dd^D p_2 \dd^D p_3 \del^D(p_1 + p_2 + p_3) \nonumber\\
&\times \left[  (p_1 - p_2)^\gamma \eta^{\mu\beta} + (p_2 - p_3)^\mu \eta^{\beta\gamma} + (p_3 - p_1)^\beta \eta^{\gamma\mu} \right] \label{eq:H1general} \\
&\times \left[  (p_1 - p_2)^{\gamma'} \eta^{\mu'\beta'} + (p_2 - p_3)^{\mu'} \eta^{\beta'\gamma'} + (p_3 - p_1)^{\beta'} \eta^{\gamma'\mu'} \right]
H^{(0)}_{\beta\beta'}(p_2) H^{(0)}_{\gamma\gamma'}(p_3) . \nonumber
\end{align}
Notice that the basic structure of the perturbative calculation is that of gauge theory. The double copy upgrades the gauge-theoretic perturbation into a calculation appropriate for gravity, coupled to a dilaton and an antisymmetric tensor. \\

As a simple example of this formalism at work, let us compute the first order correction to the simple fat graviton eq.~\eqref{eq:fatJNW} corresponding to a metric and scalar field. To begin, we need to write $H^{(0)}_{\mu\nu}(p)$ in momentum space; it is simply
\begin{equation}
H^{(0)\mu\nu}(p) = \frac{\kappa}{2} M u^\mu u^\nu \frac{\del^1(p^0)}{p^2} .
\end{equation}
Inserting this into our expression for $H^{(1)}$, eq.~\eqref{eq:H1general},
we quickly find
\begin{equation}
H^{(1)\mu\mu'}(-p_1) = \left(\frac{\kappa}{2} \right)^2 \frac{M^2}{4p_1^2} \
\int\!\dd^3 p_2\:\!\dd^3 p_3\:\!\del^4(p_1 + p_2 + p_3)
\frac{(p_2-p_3)^\mu (p_2-p_3)^{\mu'}}{p_2^2 \; p_3^2}.
\end{equation}
where $p_2^0 = 0 = p_3^0$, and consequently $p_1^0 = 0$. For future use, we note that $p_{1\mu} H^{(1)\mu\mu'}(-p_1) = 0$. Since all of the components of $H^{(1)}$ in the time direction vanish, we need only calculate the spatial components $H^{(1)ij}$. To do so, it is convenient to Fourier transform back to position space and compute firstly the Laplacian of $\nabla^2 H^{(1)ij}(x)$;
we find
\begin{align}
   \nabla^2 H^{(1)ij}\!&
    = -\left(\frac{\kappa}{2}\right)^{2} \frac{M^2}{4}\!
      \int\!\dd^3 p_2\:\!\dd^3 p_3
      \frac{ e^{-i\mathbf{p}_2 \cdot \mathbf{x}}
             e^{-i\mathbf{p}_3 \cdot \mathbf{x}} }
           { \mathbf{p}_2^2 \mathbf{p}_3^2 }
      (\mathbf{p}_2-\mathbf{p}_3)^i (\mathbf{p}_2-\mathbf{p}_3)^j  \nonumber \\ &
    = \left(\frac{\kappa}{2}\right)^{2} \frac{M^2}{4}\!
      \int\!d^3 y\:\!\delta^{(3)}(\mathbf{x}-\mathbf{y})
      (\nabla_\mathbf{x}^i - \nabla_\mathbf{y}^i)
      (\nabla_\mathbf{x}^j - \nabla_\mathbf{y}^j)
      \frac{1}{4\pi|\mathbf{x}|}
      \frac{1}{4\pi|\mathbf{y}|} \nonumber \\ &
   = -\left(\frac{\kappa}{2}\right)^{2} \frac{M^2}{4(4\pi)^2}
      \bigg( \frac{2\delta^{ij}}{r^4}
           - \frac{4 x^i x^j}{r^6}
      \bigg) .
\label{fatgravitonsimplest1laplacian}
\end{align} 
It is now straightforward to integrate this expression using spherical symmetry and the known boundary conditions to find
\begin{equation}
H^{(1)}_{\mu\nu}(x) = -\left(\frac{\kappa}{2}\right)^2 \frac{M^2}{4(4\pi r)^2} \hat{r}_\mu \hat{r}_\nu,
\label{eq:fatJNWH1}
\end{equation}
where $\hat{r}_\mu = (0, \mathbf{x}/r)$.  \\

It is interesting to pause for a moment to contrast this calculation with its analogue in Yang-Mills theory. The simplest gauge counterpart of the JNW linearised fat graviton is
\begin{equation}
A^{(0)a}_\mu(x) = g c^a u_\mu \frac{1}{4\pi r}
\qquad \Rightarrow \qquad
A^{(0)a}_\mu(p) = g c^a u_\mu \frac{\del^1(p^0)}{p^2}.
\end{equation}
To what extent is the first non-linear correction to the Yang-Mills equation similar to the equivalent in our double-copy theory? The answer to this question is clear: they are distinctly different. Indeed, the colour structure of $A^{(1)a}_\mu$ is $f^{abc} c^b c^c =0$, so $A^{(1)a}_\mu = 0$. However, the kinematic numerator of $A^{(1)a}_\mu$ identified by colour-kinematics duality is non-zero, so there is no reason for $H^{(1)}_{\mu\nu}$ to vanish. How the double copy propagates physical information from one theory to the other is unclear, but as a mathematical statement there is no issue with using the double copy to simplify gravitational calculations. \\

Given our expression, eq.~\eqref{eq:fatJNWH1}, for the fat graviton, it is now straightforward to extract the trace and the symmetric fields:
\begin{align}
\tilde \phi^{(1)} &\equiv H^{(1)}
= -\left(\frac{\kappa}{2}\right)^2 \frac{M^2}{4(4\pi r)^2} , \\
\tilde \h^{(1)}_{\mu\nu} &\equiv
\frac12 \left(H^{(1)}_{\mu\nu} + H^{(1)}_{\nu\mu} \right)
= - \left(\frac{\kappa}{2}\right)^2 \frac{M^2}{4(4\pi r)^2} \hat{r}_\mu \hat{r}_\nu .
\end{align}
However, we cannot directly deduce that this $\tilde \phi^{(1)}$ is the usual dilaton and that $\tilde \h^{(1)}_{\mu\nu}$ is the first order correction to the metric in some well-known gauge. The double copy is only guaranteed to compute quantities which are field redefinitions or gauge transformations of the graviton and dilaton. This suggests structuring calculations to compute only quantities which are invariant under field redefinitions and gauge transformations~\cite{BjerrumBohr:2004mz,Bjerrum-Bohr:2013bxa,Bjerrum-Bohr:2014zsa,Bjerrum-Bohr:2016hpa,Porto:2016pyg,Goldberger:2016iau}. However, if desired, it is nevertheless possible to determine explicitly the relevant field redefinitions and gauge transformations. This is the topic of the next section.

\subsection{Relating fat and skinny fields: gauge transformations and field redefinitions}

In section~\ref{sec:linear}, we argued that the relationship between the fat and skinny fields in linear theory is
\begin{equation}
H^{(0)}_{\mu\nu}(x) = \h^{(0)}_{\mu\nu}(x) + B^{(0)}_{\mu\nu}(x) + P^q_{\mu\nu} (\phi^{(0)}(x) - \h^{(0)}(x)).
\end{equation}
Beyond linear theory, we can expect perturbative corrections to this formula, so that
\begin{equation}
H^{}_{\mu\nu}(x) = \h^{}_{\mu\nu}(x) + B^{}_{\mu\nu}(x) + P^q_{\mu\nu} (\phi^{}(x) - \h^{}(x)) + \mathcal{O}(\kappa).
\end{equation}
We define a quantity $\mathcal{T}_{\mu\nu}$, which we call the \emph{transformation function} to make this equation exact:
\begin{equation}
H^{(1)}_{\mu\nu}(x) = \h^{(1)}_{\mu\nu}(x) + B^{(1)}_{\mu\nu}(x) + P^q_{\mu\nu} (\phi^{(1)}(x) - \h^{(1)}(x)) + \mathcal{T}^{(1)}_{\mu\nu}.
\end{equation}
We can require that $\mathcal{T}^{(1)}_{\mu\nu}$ is only constructed from linearised fields, so that $\mathcal{T}^{(1)}_{\mu\nu} = \mathcal{T}^{(1)}_{\mu\nu}(\h^{(0)}_{\alpha\beta}, B^{(0)}_{\alpha\beta}, \phi^{(0)}).$ More generally, at the $n$th order of perturbation theory 
\begin{equation}
H^{(n)}_{\mu\nu}(x) = \h^{(n)}_{\mu\nu}(x) + B^{(n)}_{\mu\nu}(x) + P^q_{\mu\nu} (\phi^{(n)}(x) - \h^{(n)}(x)) + \mathcal{T}^{(n)}_{\mu\nu}(\h^{(m)}_{\alpha\beta}, B^{(m)}_{\alpha\beta}, \phi^{(m)}),
\label{Tmunun}
\end{equation}
where $m < n$. We can therefore determine $\mathcal{T}^{(n)}_{\mu\nu}$ iteratively in perturbation theory. \\

Before we compute $\mathcal{T}^{(1)}_{\mu\nu}$ explicitly, let us pause for a moment to discuss its physical significance. Our understanding of $\mathcal{T}^{(n)}_{\mu\nu}$ rests on two facts. Firstly, the double copy is known to work to all orders in perturbation theory for tree amplitudes. Secondly, the classical background field which we have been discussing is a generating function for tree scattering amplitudes. Therefore it must be the case that scattering amplitudes computed from the classical fat graviton background fields equal their known expressions. So consider computing $H^{(n)}_{\mu\nu}$ via the double copy, and computing $\h^{(n)}_{\mu\nu}, B^{(n)}_{\mu\nu}$ and $\phi^{(n)}$ using a standard perturbative solution of their coupled equations of motion. Then the difference $H^{(n)}_{\mu\nu} - \h^{(n)}_{\mu\nu} - B^{(n)}_{\mu\nu}(x) - P^q_{\mu\nu} (\phi^{(n)}(x) - \h^{(n)}(x) )\equiv \mathcal{T}^{(n)}_{\mu\nu}$ must vanish upon use of the LSZ procedure. We conclude that $\mathcal{T}_{\mu\nu}$ parameterises redundancies of the physical fields which are irrelevant for computing scattering amplitudes: gauge transformations and field redefinitions. Indeed, the very definition of $\mathcal{T}_{\mu\nu}$ requires choices of gauge: for example, the choice of de Donder gauge for the skinny graviton. \\

Since $\mathcal{T}_{\mu\nu}$ parameterises choices which can be made during a calculation, such as the choice of gauge, we do not expect a particularly simple form for it. Nevertheless, to compare explicit skinny gravitons computed via the double copy with standard metrics, it may be useful to have an explicit form of $\mathcal{T}^{(1)}_{\mu\nu}$. It is always possible to compute $\mathcal{T}^{(n)}_{\mu\nu}$ directly through its definition, at the expense of perturbatively solving the coupled Einstein, scalar and antisymmetric tensor equations of motion. For example, consider the fat graviton $H^{(1)}_{\mu\nu}(x)$, eq.~\eqref{eq:fatJNWH1}, we computed in the previous section. Since there is no antisymmetric tensor in this system, we may compute $\mathcal{T}^{(1)}_{\mu\nu}$ under the simplifying assumption that $B_{\mu\nu} = 0$ so that $H_{\mu\nu}$ is symmetric. We find that when
$\partial_\mu \h^{(0)\mu\nu} = \partial_\mu H^{(0)\mu\nu} = 0$,
then the transformation function is
\begin{equation}
\begin{split}
\mathcal{T}^{(1)\mu\nu}(-p_1)
 = \int& \dd^Dp_2\dd^Dp_3 \del^{D}(p_1+p_2+p_3)
\frac{1}{4p_1^2}
\bigg\{
   H_{2\alpha\beta}^{(0)} H_3^{(0)\alpha\beta} p_1^{\mu}p_1^{\nu}
 + 8 p_{2}^{\alpha} H^{(0)}_{3\alpha\beta}
     H^{(0)\beta(\mu}_{2} p_1^{\nu)} \\ &
 + 8 p_2 \cdot p_3\,H_2^{(0)\mu\alpha} H^{(0)\nu}_{3~~\alpha}
 - 2\eta^{\mu\nu} p_2\cdot p_3\,H_{2\alpha\beta}^{(0)} H_3^{(0)\alpha\beta}
 + 4\eta^{\mu\nu} p_2^{\alpha} H^{(0)}_{3\alpha\beta}
   H^{(0)\beta\gamma}_{2} p_{3\gamma} \\ &
 + P_q^{\mu\nu} \left[
   2(D-6) p_2\cdot p_3\,H_{2\alpha\beta}^{(0)} H_3^{(0)\alpha\beta}
 - 4(D-2) p_2^{\alpha} H^{(0)}_{3\alpha\beta}
   H^{(0)\beta\gamma}_{2} p_{3\gamma} \right]
\bigg\},
\end{split}
\end{equation}
where we have used a convenient short-hand notation
\begin{align}
H_i^{\mu\nu} &\equiv H^{\mu\nu}(p_i),  \qquad \quad
p^{(\mu} q^{\nu)} \equiv \frac12 \left( p^\mu q^\nu + p^\nu q^\mu \right).
\end{align}
This expression is valid for any symmetric $H_{\mu\nu}^{(0)}$, and the extension to general $H_{\mu\nu}^{(0)}$ is straightforward. \\

While the information in the transformation function contains little content of physical interest, it may be of some interest from the point of view of the mathematics of colour-kinematics duality. Indeed, in the special case of the self-dual theory, it is known how to choose an explicit parameterisation of the metric perturbation so that the double copy is manifest~\cite{Monteiro:2011pc}. Choosing these variables therefore sets $\mathcal{T}_{\mu\nu} = 0$ to all orders, for self-dual spacetimes. Once the relevant variables have been chosen, then the kinematic algebra in the self-dual case was manifest at the level of the equation of motion of self-dual gravity: the algebra is one of area-preserving diffeomorphisms. Perhaps it is the case that an understanding of the transformation function in the general case will open the way towards a simple understanding of the full kinematic algebra.

\subsection{The perturbative corrections to the JNW fields}

We are now in a position to convert our fat graviton $H^{(1)}_{\mu\nu}(x)$, eq.~\eqref{eq:fatJNWH1} into skinny fields. The simple form of the $H^{(0)}_{\mu\nu}(x)$ leads to a simplification in the transformation function, since $p\cdot u =0$ for a stationary source. Thus $\mathcal{T}^{(1)\mu\nu}$ is simply
\begin{equation}
\begin{split}
\mathcal{T}^{(1)\mu\nu}(-p_1)= -\left( \frac{\kappa}{2} \right)^2 M^2 \int& \dd^4p_2\dd^4p_3 \del^{4}(p_1+p_2+p_3)
\frac{1}{4p_1^2}
\frac{\del^{1}(p_2^0)}{p_2^2}\frac{\del^{1}(p_3^0)}{p_3^2} \\ 
\times&\bigg\{ 8p_2\cdot p_3 u^\mu u^\nu -p_1^{\mu}p_1^{\nu}+2\eta^{\mu\nu}p_2\cdot p_3+P_q^{\mu\nu}\left[4p_2\cdot p_3\right]
\bigg\},
\end{split}
\end{equation}
in $D=4$. Performing the Fourier transform, we find
\begin{equation}
\mathcal{T}^{(1)}_{\mu\nu}(x) = - \left(\frac{\kappa}{2}\right)^{2} 
 \left[3u_\mu u_\nu +2 \hat{r}_\mu \hat{r}_\nu+2P^q_{\mu\nu}\right]\frac{M^2}{4 (4\pi r)^2}.
\end{equation}
Let us now extract the skinny fields in de Donder gauge from our fat graviton, eq.~\eqref{eq:fatJNWH1}. The relation between the fat and skinny fields is now given by
\begin{align}
\h^{(1)}_{\mu\nu}(x) + P^q_{\mu\nu} \left[\phi^{(1)}(x) - \h^{(1)}(x) \right] &= H^{(1)}_{\mu\nu}(x) -\mathcal{T}^{(1)}_{\mu\nu}(x) \\
&= - \left(\frac{\kappa}{2}\right)^{2} \hat{r}_\mu \hat{r}_\nu \frac{M^2}{4(4\pi r)^2} + \left(\frac{\kappa}{2}\right)^{2} 
 \left[3u_\mu u_\nu +2 \hat{r}_\mu \hat{r}_\nu+2P^q_{\mu\nu}\right]\frac{M^2}{4 (4\pi r)^2}.
\nonumber
\end{align}
Thus, the dilaton vanishes as anticipated in section~\ref{sec:linearJNW}, since
\begin{align}
\phi^{(1)}(x) &= H^{(1)}(x) - \mathcal{T}^{(1)}(x) = 0.
\end{align}
Consequently, the negative of the trace of the metric is the only term acted upon by $P_q^{\mu\nu}$, so we find
\begin{equation}
\h^{(1)}(x) = -\left(\frac{\kappa}{2}\right)^{2} \frac{M^2}{2(4\pi r)^2},
\end{equation}
The metric is easily seen to be
\begin{equation}
\h^{(1)}_{\mu\nu}(x) = \left(\frac{\kappa}{2}\right)^{2} 
 \left(3 u_\mu u_\nu + \hat{r}_\mu \hat{r}_\nu \right) \frac{M^2}{4(4\pi r)^2},
\end{equation}
consistent with the anticipated trace, and in agreement with the known result for the JNW metric, eq.~\eqref{eq:JNWexpansion}, when $M=Y$.

\subsection{Higher orders}
\label{sec:higher}

In section~\ref{sec:firstNLO}, we saw how fat graviton fields can be obtained
straightforwardly from perturbative solutions of the Yang-Mills
equations. These can then be
translated to skinny fields, if necessary, after obtaining the
relevant transformation functions ${\cal T}^{\mu\nu}$. Now let us
briefly describe how this procedure generalises to higher orders.\\

As we explained in section~\ref{sec:review}, the validity of the double copy 
relies on writing Yang-Mills diagrams such that colour-kinematics duality is
satisfied.
But, in general, a perturbative solution of the conventional
Yang-Mills equations will not satisfy this property. So before using the double copy, one 
must reorganise the perturbative
solution of the theory so that, firstly, only three-point interaction vertices
between fields occur, and secondly, the numerators of these three-point diagrams
satisfy the same algebraic identities (Jacobi relations and antisymmetry properties) as the colour factors. 
The Jacobi identities can be enforced by using an explicit Yang-Mills Lagrangian designed for this purpose~\cite{Bern:2010yg,Tolotti:2013caa}. It is known how to construct this Lagrangian to arbitrary order in perturbation theory. This Lagrangian is non-local and contains Feynman vertices with an infinite number of fields. If desired, it is possible to obtain a local Lagrangian containing only three point vertices at the expense of introducing auxiliary fields. 
For now, we will restrict ourselves to four-point order. 
At this order Bern, Dennen, Huang and Kiermaier (BDHK) introduced~\cite{Bern:2010yg} an auxiliary field $B^a_{\mu\nu\rho}$ so as to write a cubic version of the Yang-Mills Lagrangian,
\begin{equation}
\label{eq:BDHK}
\mathcal{L}_\textrm{BDHK}
 = \frac12 A^{a\mu} \partial^2 A^a_\mu
 + B^{a\mu\nu\rho} \partial^2 B^a_{\mu\nu\rho}
 - g f^{abc} \left(\partial_\mu A^a_\nu
 - \partial^\rho B^a_{\rho\mu\nu}\right) A^{b\mu} A^{c\nu} .
\end{equation}
Since the role of the field $B^a_{\mu\nu\rho}$ is essentially to be a Lagrange multiplier, it is understood that no sources for $B^a_{\mu\nu\rho}$ should be introduced. \\

To illustrate the procedure in a non-trivial example, let us compute the second order correction to the JNW fat graviton, $H^{(2)}_{\mu\nu}(x)$. In fact, a number of simplifications make this calculation remarkably straightforward. Firstly, the momentum space equation of motion for the auxiliary field appearing in the BDHK Lagrangrian, eq.~\eqref{eq:BDHK}, is
\begin{equation}
p_1^2 B^{(1)a}_{\mu\nu\rho}(-p_1) = \frac{i}{4} f^{abc} \int \dd^4 p_2 \dd^4 p_3 \del^4(p_1+p_2+p_3) p_{1\mu} \left[\eta_{\nu \beta}\eta_{\rho \gamma}  - \eta_{\nu \gamma} \eta_{\rho\beta} \right] A^{(0)b\beta}(p_2) A^{(0)c\gamma}(p_3).
\end{equation}
Notice that the term in square brackets is antisymmetric under interchange of $\beta$ and $\gamma$; imposing this symmetry is a requirement of colour-kinematics duality because the associated colour structure is antisymmetric under interchange of $b$ and $c$. 
A consequence of this simple fact is that, in the double copy, the auxiliary field vanishes in the JNW case (to this order of perturbation theory).
In fact, two auxiliary fields appear in the double copy: one can take two copies of the field $B$, or one copy of $B$ times one copy of the gauge boson $A$. In either case, the expression for an auxiliary field in the double copy in momentum space will contain a factor
\begin{align}
p_{1\mu} \left[\eta_{\nu \beta}\eta_{\rho \gamma}  - \eta_{\nu \gamma} \eta_{\rho\beta} \right] H^{(0)\beta\beta'}(p_2) H^{(0)\gamma\gamma'}(p_3) &= 
p_{1\mu} \left[\eta_{\nu \beta}\eta_{\rho \gamma}  - \eta_{\nu \gamma} \eta_{\rho\beta} \right] \frac{\delta^1(p_2^0)}{p_2^2}\frac{\delta^1(p_3^0)}{p_3^2} u^\beta u^{\beta'} u^\gamma u^{\gamma'}
= 0,
\end{align}
because of the antisymmetry of the vertex in square brackets, and the factorisability of the tensor structure of the zeroth order JNW expression. \\

Consequently, the Yang-Mills four-point vertex plays no role in the the double copy for JNW at second order. Thus the Yang-Mills equation to be solved is simply
\begin{multline}
p_1^2 A^{(2)a\mu}(-p_1) = i f^{abc} \int \dd^4 p_2 \dd p_3 \del^4(p_1 + p_2 + p_3) \\
\times \left[  (p_1 - p_2)^\gamma \eta^{\mu\beta} + (p_2 - p_3)^\mu \eta^{\beta\gamma} + (p_3 - p_1)^\beta \eta^{\gamma\mu} \right]  A^{(0)b}_\beta(p_2) A^{(1)c}_\gamma(p_3) ,
\label{eq:NLOcorrectionYM}
\end{multline}
using the symmetry of the expression under interchange of $p_2$ and $p_3$. Thus, $H^{(2)}$ is the solution of
\begin{align}
p_1^2 H^{(2)\mu\mu'}(-p_1) = &\,\frac{1}{2} \int \dd^4 p_2 \dd^4 p_3 \del^4(p_1 + p_2 + p_3) \nonumber \\
&\times \left[  (p_1 - p_2)^\gamma \eta^{\mu\beta} + (p_2 - p_3)^\mu \eta^{\beta\gamma} + (p_3 - p_1)^\beta \eta^{\gamma\mu} \right] \label{eq:H2} \\
&\times \left[  (p_1 - p_2)^{\gamma'} \eta^{\mu'\beta'} + (p_2 - p_3)^{\mu'} \eta^{\beta'\gamma'} + (p_3 - p_1)^{\beta'} \eta^{\gamma'\mu'} \right]
H^{(0)}_{\beta\beta'}(p_2) H^{(1)}_{\gamma\gamma'}(p_3) . \nonumber
\end{align}
This expression simplifies dramatically when we recall that $H^{(0)}_{\beta\beta'}(p_2)$ and $H^{(1)}_{\gamma\gamma'}(p_3)$ both have vanishing components of momentum in the time direction, so that $p_2^0 = 0 = p_3^0 = p_1^0$. Meanwhile $H^{(0)}_{\beta\beta'}(p_2) \propto u_\beta u_{\beta'}$. Thus,
\begin{align}
p_1^2 H^{(2)}_{\mu\mu'}(-p_1)
= 2 \int \dd^4 p_2 \dd^4 p_3 \del^4(p_1 + p_2 + p_3)
H^{(0)}_{\mu\mu'}(p_2)\,p_{2}^{\alpha} H^{(1)}_{\alpha\beta}(p_3) p_2^{\beta}.
\end{align}
We find it convenient to Fourier transform back to position space, where we must solve the simple differential equation
\begin{equation}
\partial^2 H^{(2)}_{\mu\mu'}(x) = 2 H^{(1)}_{\alpha \alpha'} \partial^\alpha \partial^{\alpha'} H^{(0)}_{\mu \mu'}.
\end{equation}
Inserting explicit expressions for $H^{(0)}$, eq.~\eqref{eq:fatJNW} and $H^{(1)}$, eq.~\eqref{eq:fatJNWH1}, and bearing in mind that the situation is static, the differential equation simplifies to
\begin{equation}
\nabla^2 H^{(2)}_{\mu\mu'}(x) = -\left( \frac\kappa2\right)^3 \frac{M^3}{(4\pi r)^3} \frac{u_\mu u_{\mu'}}{r^2},
\end{equation}
with solution
\begin{equation}
H^{(2)}_{\mu\mu'}(x) = -\left( \frac\kappa2\right)^3 \frac{M^3}{6 (4\pi r)^3} u_\mu u_{\mu'}.
\end{equation}
We could now, if we wished, extract the metric perturbation and scalar 
field corresponding to this expression. Indeed, it is always possible to convert fat gravitons into
ordinary metric perturbations in a specified gauge.  \\

It is possible to continue to continue this calculation to higher orders. In that case, more work is required in order to satisfy the requirement of colour-kinematics duality. It is possible to
supplement the BDHK Lagrangian by higher-order effective operators
involving the gluon field, constructed order-by-order in perturbation
theory, which act to enforce colour-kinematics duality. Furthermore, one may introduce further auxiliary fields
so that only cubic interaction terms appear in the Lagrangian. This procedure
is explained in detail in refs.~\cite{Bern:2010yg,Tolotti:2013caa}, and can be carried out to arbitrary perturbative order. The
fat graviton equation of motion is constructed as a term-by-term double copy
of the fields in the colour-kinematics satisfying Yang-Mills Lagrangian.
In this way, it is possible to calculate perturbative fat
gravitons to any order using Yang-Mills theory and the double copy.

\section{Discussion}
\label{sec:discuss}

In this paper, we have addressed how classical solutions of
gravitational theories can be obtained by double-copying Yang-Mills
solutions. These results go beyond the classical double copies of
refs.~\cite{Monteiro:2014cda,Luna:2015paa,Luna:2016due,Ridgway:2015fdl,Borsten:2013bp,Anastasiou:2013hba,Anastasiou:2014qba,Anastasiou:2016csv,Cardoso:2016ngt,Cardoso:2016amd}
in that the solutions are non-linear. However, the price one pays is
that they are no longer exact, but must be constructed order-by-order
in perturbation theory. We have concentrated on solutions obtained
from two copies of pure (non-supersymmetric) Yang-Mills theory, for
which the corresponding gravity theory is ${\cal N}=0$
supergravity. The double copy then relates the Yang-Mills fields to a
single field $H_{\mu\nu}$, that we call the {\it fat graviton}, and
which in principle can be decomposed into its constituent {\it
  skinny fields}, which we take to be the graviton $\h^{\mu\nu}$
(defined according to eq.~(\ref{gothich})), the dilaton $\phi$, and
the two-form $B^{\mu\nu}$. \\

Our procedure for calculating gravity solutions is as follows:
\begin{enumerate}
\item For a given distribution of charges, one may perturbatively
  solve the Yang-Mills equations for the gauge field $A^{\mu\,a}$,
  given in terms of integrals of interaction vertices and propagators.
\item The solution for the fat graviton is given by double copying the
  gauge theory solution expression according to the rules of
  refs.~\cite{Bern:2008qj,Bern:2010ue,Bern:2010yg} once colour-kinematics duality is satisfied. That is, one
  strips off all colour information, and duplicates the interaction
  vertices, leaving propagators intact.
\item The fat graviton can in principle be translated into skinny
  fields using the transformation law of eq.~(\ref{Tmunun}), which
  iteratively defines the {\it transformation function} ${\cal
    T}^{\mu\nu}$. This function can be obtained from matching the fat
  graviton solution to a perturbative solution of the conventional
  ${\cal N}=0$ supergravity equations. Once found, however, it can be
  used for arbitrary source distributions.
\end{enumerate}
The presence of the transformation function ${\cal T}^{\mu\nu}$ is at
first glance surprising. One may always decompose the fat graviton in
terms of its symmetric traceless, anti-symmetric and trace degrees of
freedom. Then one could simply define that these correspond to the
physical graviton, two-form and dilaton. However, one has the freedom
to perform further field redefinitions and gauge transformations of
the skinny fields, in order to put these into a more conventional
gauge choice (e.g. de Donder). The role of ${\cal T}^{\mu\nu}$ is then
to perform this redefinition. It follows that it carries no physical
degrees of freedom itself, and indeed is irrelevant for any physical
observable. \\

We have given explicit examples of fat gravitons, and their relation
to de Donder gauge skinny fields, up to the first subleading order in
perturbation theory. We took a stationary point charge as our source,
finding that one can construct either the Schwarzschild metric (as in
the Kerr-Schild double copy of ref.~\cite{Monteiro:2014cda}), or the
JNW solution~\cite{Janis:1968zz} for a black hole with non-zero scalar
field $\phi$. Which solution one obtains on the gravity side amounts
to the choice of whether or not to source the dilaton upon performing
the double copy. This mirrors the well-known situation for amplitudes,
namely that the choice of polarisation states in gauge theory
amplitudes determines whether or not a dilaton or two-form is obtained
in the corresponding gravity amplitudes at tree level. This clarifies
the apparent puzzle presented in ref.~\cite{Goldberger:2016iau},
regarding whether it is possible for the same gauge theory solution to
produce different gravity solutions.
\\

Underlying the simplicity of the double copy is the mystery of the kinematic algebra.
While it is known that one can always find kinematic numerators for gauge theory
diagrams so that colour-kinematics duality is satisfied, it is not known whether an
off-shell algebraic structure exists in the general case which can compute these numerators.
If this algebra exists, it may further simplify the calculations we have described in
this paper. The kinematic algebra should allow for a more algebraic computation of
the numerators of appropriate gauge-theoretic diagrams, perhaps without the need
for auxiliary fields. Similarly, it seems possible that a detailed understanding of the
kinematic algebra will go hand-in-hand with deeper insight into the transformation
function $\mathcal{T}_{\mu\nu}$ which parameterises the choice of gauge and
field redefinition picked out by the double copy. \\

Our ultimate aim is to use the procedure outlined in this paper in
astrophysical applications, namely to calculate gravitational
observables for relevant physical sources (a motivation shared by
ref.~\cite{Goldberger:2016iau}). To this end, our fat graviton
calculations must be extended to include different sources, and also
higher orders in perturbation theory. In order to translate the fat
graviton to more conventional skinny fields, one would then need to
calculate the relevant transformation functions ${\cal T}^{(n)}_{\mu\nu}$. An alternative possibility exists, namely to
calculate physical observables, which must be manifestly invariant
under gauge transformations and field redefinitions, directly from fat
graviton fields, without referring to skinny fields at all. Work on
these issues is ongoing.

\section*{Acknowledgements}

We are very grateful to Alex Anastasiou and Radu Roiban for
useful discussions.
We especially thank John Joseph Carrasco for emphasising the relevance of the fat graviton (and providing its name).
DOC is supported in part by
the STFC consolidated grant Particle Physics at the Higgs Centre,
while DOC and AO are supported in part by the Marie Curie FP7 grant 631370.
IN is supported by an STFC studentship.
NW acknowledges support from the EPSRC CM-CDT Grant No. EP/L015110/1.
CDW is supported by the UK Science and Technology Facilities Council (STFC),
and AL by a Conacyt studentship and a Lord Kelvin Fund travel scholarship.
Both CDW and AL thank the Higgs Centre for Theoretical Physics, University of Edinburgh, for warm hospitality.
RM, DOC, AO and CDW are grateful to Nordita for hospitality during
the programme ``Aspects of Amplitudes.''
RM, DOC, AL, AO and CDW would like to thank the Isaac Newton Institute
for Mathematical Sciences for its hospitality during
the programme ``Gravity, Twistors and Amplitudes''
which was supported by EPSRC Grant No. EP/K032208/1.

\bibliography{refs.bib}
\end{document}